\documentclass[prl,twocolumn,aps]{revtex4}
\usepackage{graphicx}
\usepackage{latexsym}
\usepackage{amsfonts}
\usepackage{amsmath}
\usepackage{amssymb}
\usepackage{revsymb}
\usepackage{bm}
\usepackage{color}
\usepackage{cancel}
\usepackage{feynmf}
\usepackage{graphicx}
\usepackage{epstopdf}
\usepackage{pstricks}
\begin{document}

\title{Cold Dark Matter with MOND Scaling}
%\title{Holographic Dark Matter}
\author{Chiu Man Ho}\email{chiuman.ho@vanderbilt.edu}
\author{Djordje Minic}\email{dminic@vt.edu}
\author{Y. Jack Ng}\email{yjng@physics.unc.edu}
\affiliation{Department of Physics and Astronomy, Vanderbilt University, Nashville, TN 37235, U.S.A.}
\affiliation{IPNAS, Department of Physics, Virginia Tech, Blacksburg, VA 24061, U.S.A.}
\affiliation{Institute of Field Physics, Department of Physics and Astronomy,
University of North Carolina, Chapel Hill, NC 27599, U.S.A.}

%\rightline{VPI-IPNAS-09-04}

%\renewcommand{\thefootnote}{\fnsymbol{footnote}}
\date{\today}
\begin{abstract}
We provide a holographic dual description of Milgrom's scaling
associated with galactic rotation curves. Our argument is partly based on the
recent entropic reinterpretation of Newton's laws of motion.
We propose a duality between cold dark matter and modified Newtonian dynamics (MOND).
We introduce the concept of MONDian dark matter, and discuss some of its phenomenological implications.
At cluster as well as cosmological scales, the MONDian dark matter would behave as cold dark matter,
but at the galactic scale, the MONDian dark matter would act as MOND.

\end{abstract}

\maketitle

%%%%%%%%%%%%%%%%%%%%%%%%%%%%%%%%%%%%%%%%%%%%%%%%%%%%%%%%%%%%%%%%%%%%%%%%%%%%%
%\section*{Outline of the paper}
%%%%%%%%%%%%%%%%%%%%%%%%%%%%%%%%%%%%%%%%%%%%%%%%%%%%%%%%%%%%%%%%%%%%%%%%%%%%%%%%%

\emph{Introduction.} One of the most outstanding puzzles of contemporary physics is the
nature of  the ``missing mass'' or dark matter \cite{dark}.
That cold dark matter (CDM) should exist is strongly supported by various
observations such as the galactic rotation
curves, the large scale
structure surveys and the cosmic microwave background \cite{dark}. On the other hand, there
is a remarkable observation due to Milgrom, regarding
a very successful scaling observed in the galactic rotation curves
that goes by the name of modified Newtonian dynamics, or MOND \cite{mond}.
Milgrom's approach aims to reinterpret the ``missing mass''
problem as the ``acceleration discrepancy'', and thus points
to a radical modification of gravity
and the laws of motion \cite{teves}. Specifically, Milgrom postulates that the
acceleration of a test mass $m$ due to the source $M$ is given by
\begin{eqnarray}
a= \left\{
     \begin{array}{ll}
       a_N
       %= \frac{G \,M }{r^2}
       & \hbox{~~~$a \gg a_c$} \\
       \sqrt{a_N\, a_c}
       %= \frac{\sqrt{G\, M \,a_c}}{r}
       & \hbox{~~~$a \ll a_c$}
     \end{array}
   \right.
\end{eqnarray}
where $a_N= G M /r^2$ is the magnitude of
the usual Newtonian acceleration. Coincidentally,
the critical acceleration $a_c$ is related to the speed of light $c$ and the Hubble scale $H$:
$
a_c \sim c H/(2 \pi) \sim 1.2 \times 10^{-8} cm/s^2.
$
It turns out that MOND fits hundreds of galactic rotation
curves very well and the Tully-Fisher relation is automatically satisfied. Since the galactic dynamics is
very complex, it is not surprising that MOND doesn't explain all of the observed
galactic rotation curves. But the success of this simple MOND relation seems to suggest that it may hold the
key to the problem.

The basic difference between the above two approaches is nicely encapsulated
in two possible modifications of Einstein's equations of motion
$
G_{\mu\nu} \equiv R_{\mu\nu} - \frac{1}{2} g_{\mu\nu} R = \Lambda g_{\mu\nu} + 8 \pi G_N T_{\mu\nu}.
$
One can either change the source term,
%the energy momentum tensor
$T_{\mu\nu} \to T_{\mu\nu} + T_{\mu\nu}^{DM}$, which is the conventional
dark matter approach, or one can attempt to change
the Einstein tensor $G_{\mu\nu} \to F(G_{\mu\nu})$ (where $F$ is
either a local or possibly non-local operator), and thus modify
gravity, which is the approach associated with MOND.

%While Milgrom's scaling has been phenomenologically very successful
%(and to some extent, superior to CDM)
%at the galactic scale \cite{dsmond},
While the CDM paradigm has its attractive features, it cannot easily explain the
observed
galactic flat rotation curves and the observed Tully-Fisher relation \cite{dsmond} that the
MOND scenarios can. On the other hand,
there are problems with MOND at the cluster and cosmological scales,
where apparently CDM works much better \cite{dark}.
This inspires us to ask: Could there be some kind of dark matter
that can behave like MOND at the galactic scale?

The two ideas, one based on the existence of dark matter
and one that denies it, but requires the radical modification
of the laws of motion, are apparently contradicting and hence irreconcilable.
However, that seemingly incompatible ideas can be incorporated in
a new concept is well known in the history of physics,
with the wave-particle duality being one of the most
astonishing instances. Inspired by such lofty examples,
%in this Letter,
we would like to suggest a reconciliation of the dark matter
and MOND approaches
by introducing a new concept of ``MONDian dark matter".
In a nutshell, {\it we propose a scheme such that the MONDian dark matter
looks like CDM at cluster and cosmological
scales, but it behaves like MOND at the galactic scale.}
This would perhaps explain
%the current difficulties in observing the
%CDM particles and also
the apparent failure of MOND at
cluster and cosmological scales.
%Our approach will also
%suggest new ways of searching for MONDian dark matter.

In fact, a preliminary attempt to mimic Milgrom's scaling
from the CDM paradigm has been conducted in \cite{turner}, although it has completely ignored
the low-surface-brightness galaxies in which the acceleration is
everywhere smaller than $a_c$.
We think that it is important
to have a deeper theoretical understanding about the connection between the nature
of dark matter
and Milgrom's scaling.
%Obviously, it would be difficult to incorporate the CDM-MOND duality suggested
%by the concept of MONDian dark matter into the standard effective
%field theory approach, but
{\it It is our intention to combine the salient successful features of both
CDM and MOND into a
unified scheme, by introducing the concept of MONDian dark matter.}
%We wish to explain the MOND
%scaling, which works so well on galactic scales, starting from the CDM side!}
Hopefully our proposal could also point to quantum gravitational
origins of the ``missing mass''.

As a remark, it is not clear how MOND follows from a
relativistic modification of Einstein's gravity. One possibility is to modify
the effective metric
\cite{teves} by introducing
other degrees of freedom
$
g_{\mu\nu} \to {\tilde{g}}_{\mu\nu} (g_{\mu\nu}, A_{\mu}, \varphi...),
$
where $A_\mu$ and $\varphi$ correspond to the vector
and scalar degrees of freedom respectively.
Then, a theory is sought in terms of these degrees of
freedom, which reproduces MOND and is consistent with
some general symmetry principles (such as relativistic
invariance, causality etc). However, we are not committed to a concrete
model for such a modification of gravity.
We simply assume that the bulk space-time
gravitational theory could be deformed by these new
degrees of freedom so as to imply MOND at the galactic
scale (GS):\,
$
\int d^4 x \sqrt{-g} [ R_g + L_{SM}(\phi_g) ]\to
\int_{GS} d^4 x \sqrt{-\tilde{g}} [ R_{\tilde{g}} + L_{SM}(\phi_{\tilde{g}})],
$\,
where $L_{SM}$ is the Standard Model Lagrangian.

%The essence of Milgrom's
%proposal is that in the halos of galaxies,
%one has an effective modification of Newton's second law
%so that $a =  \sqrt{a_N a_c}$,

\emph{Entropic Reinterpretations.}
%The starting point of our holographic reinterpretation of MOND is
We start with the recent work of E. Verlinde \cite{verlinde}
in which the canonical Newton's laws are derived
from the point of view of holography.
Using the first law of thermodynamics, Verlinde \cite{verlinde} proposes the
concept of entropic force
$
F_{entropic} = T \frac{\Delta S}{\Delta x},
$
where $\Delta x$ denotes an infinitesimal spatial displacement of a particle with mass $m$ from the heat
bath with temperature $T$. He then
invokes Bekenstein's original arguments
concerning the entropy $S$ of black holes \cite{bekenstein}
%$
%S_{BH} = k_B \log \Omega,
%$
%By fixing $\delta x$ to be the Compton wavelength
%$\delta x = \frac{\hbar}{mc}$
%for the particle of given mass $m$, and
by imposing $
\Delta S = 2\pi k_B \frac{mc}{\hbar} \Delta x
$.\,
%due to a displacement
%$\Delta x = \frac{\hbar}{mc}$ associated with
%a particle that is a Compton wavelength apart.
%Verlinde postulates the entropic change linear with the
%change in distance
%$
%\Delta S = 2\pi k_B \frac{mc}{\hbar} \Delta x.
%$
Using the famous formula for the Unruh temperature,
$
k_B T = \frac{\hbar a}{ 2\pi c},
$\,
associated with a uniformly accelerating (Rindler) observer \cite{unruh},
he obtains Newton's second law
$
\vec{F} = m \vec{a},
$\,
with the vectorial form being dictated by the gradient of the entropy.

%To determine the exact expression for $a$,
Next, Verlinde considers an
imaginary quasi-local (spherical) holographic screen of area $A=4 \pi r^2$ with
temperature $T$. Then, he assumes the equipartition of energy $E= \frac{1}{2} N k_B T$ with $N$ being
the total number of degrees of freedom (bits) on the screen given by $N = Ac^3/(G \hbar)$. Using the Unruh
temperature formula and the fact that $E=M c^2$, he recovers exactly the non-relativistic Newton's law
of gravity, namely $a= G M /r^2$. This is precisely the fundamental relation that Milgrom
is proposing to modify so as to fit the galactic rotation curves.
Therefore, in view of Verlinde's proposal for the entropic \cite{bekenstein}, and thus
holographic \cite{hawking,holography,adscft}
reinterpretation of Newton's law, it is natural to ask:
What entropic or holographic interpretation lies behind
Milgrom's modification of Newton's second law?

%using the
%equipartition theorem to relate energy per
%degree of freedom and
%temperature $E= \frac{1}{2} N k_B T$
%on one side, and  on the other, where
%$M$ would be the emergent mass enclosed by the holographic screen,
%Verlinde recovers Newton's law
%of gravity $F= G_N M m/r^2$.

First, a comment on the entropic approach. While it has not yet offered any radically new physics, the
entropic
approach has consistently brought together a few crucial notions in physics and has
provided an alternative unifying point of view.  Below we will show that this approach
can be used to shed new light on dark matter.

Now to proceed,
%with this investigation,
we first have to recognize that we live in an accelerating universe.
This suggests that we will need a generalization of Verlinde's proposal to de Sitter (dS) space.
For convenience, we set $\hbar = c= 1$ henceforth.
In particular, the Unruh-Hawking temperature, as measured by an inertial observer in de Sitter space with
a positive cosmological constant $\Lambda$,
is given by $T_{dS} = \frac{1}{2\pi k_B} a_0$ where $a_0=\sqrt{\frac{\Lambda}{3}}$ \,\cite{hawking}. Notice that
$\Lambda$ is related to the Hubble scale $H$ through
$
\Lambda = 3 H^2.
$\,
The corresponding Unruh temperature as measured by a non-inertial observer with acceleration $a$ will be
\cite{deser}
\begin{equation}
T_{dS+a} = \frac{1}{2\pi k_B} \sqrt{a^2+a_0^2}\,.
\end{equation}
This formula can be derived by straightforward but lengthy
calculation. Instead, it can also be heuristically derived by noticing that
$dS^4$ can be embedded into a five-dimensional Minkowski spacetime $M^5$.
World lines with proper acceleration $a$ in $dS^4$ (parametrized by $\Lambda =3 \,a_0^2$) can be viewed as
world lines with proper acceleration $\sqrt{a^2+a_0^2}$ \,in $M^5$.
Consequently, we can define the \emph{net}
temperature as measured by the non-inertial observer
(due to some matter sources that cause the acceleration $a$\,) to be
\begin{equation}
\tilde{T}\equiv T_{dS+a}- T_{dS}
=\frac{1}{2\pi k_B} [\sqrt{a^2+a_0^2} - a_0]\,.
\end{equation}
As a remark, this formula can be formally applied to
anti-de Sitter (AdS) space as well by taking $\Lambda \to - \Lambda$.

Interestingly, Milgrom has suggested in \cite{interpol} that the difference between the
Unruh temperatures as measured by non-inertial and inertial observers in de Sitter space,
namely $2\pi k_B \Delta T =\sqrt{a^2+a_0^2} - a_0$,\,
%$
%2 \pi k_B \delta T_{dS} =  \sqrt {a^2 + \Lambda/3} -  \sqrt {\Lambda/3}
%$
could give the correct behaviors of the interpolating function between the usual Newtonian
acceleration and his suggested MONDian deformation for very small accelerations.
Even though $\sqrt{a^2+a_0^2} - a_0$ could somehow mimic the correct behaviors of his MOND theory,
Milgrom was not able to justify why the force is related to the difference between the
Unruh temperatures as measured by non-inertial and inertial observers in de Sitter space.
Or, in his own words: ``it is not really clear why $\Delta T$ should be a measure of inertia".
Thus, without a reasonable justification, his suggestion remains to be an \emph{ad hoc} mathematical function that can
reproduce the behaviors of the MOND theory. As we will see in next section, adopting
Verlinde's entropic force point of view allows us to justify Milgrom's suggestion naturally.

%Armed with Verlinde's entropic approach to Newton's second law, we can obviously expect that
%Milgrom's scaling can be derived, as a limit of the
%entropic reinterpretation of the Unruh formula, from
%the difference between the temperature associated with
%an accelerating observer in de Sitter space, and
%the ``ambient'' de Sitter temperature associated
%with the cosmological horizon.

%In particular, Verlinde's entropic force associated
%with $\delta T_{dS}$ is given by
%$
%F = \delta T_{dS} \nabla_x S = m  [\sqrt {a^2 + \Lambda/3} -  \sqrt {\Lambda/3}],
%$
%which for small accelerations reproduces Milgrom's scaling.
%We will show that this logic goes through in our case as well, if
%we start from $T_{dS}$,
%and $\delta T_{dS}$,
%except that the mass $M$ will be interpreted
%to incorporate ``MONDian dark matter'', i.e. the boundary degrees of
%freedom dual to the bulk modification of gravity, compatible with MOND.
%More specifically, when $a \gg a_c$ Milgrom proposes
%$F_{Milgrom}=m a = m a_N$, and
%when $a \ll a_c$ he proposes
%\begin{equation}
%F_{Milgrom}=m a = m \sqrt{a_N a_c}, ~~;~~ a_N= GM/r^2
%\end{equation}
%where $M$ is the mass of ordinary matter. This formula works extremely well
%at the galactic scale.

\emph{CDM-MOND Duality.}\; Following Verlinde's approach, the entropic force, acting on the test
mass $m$ with acceleration $a$ in de Sitter space, is given by
\begin{equation}
F_{entropic}=\tilde{T}\, \nabla_x S= m [\sqrt{a^2+a_0^2}-a_0].
\end{equation}
For $a \ll a_0$, we have
$
F_{entropic}\approx m \frac{a^2}{2\,a_0}.
$
%The question is: what is the exact expression for
%$a$ when $a \ll a_0$? In order to
In order to fit the galactic rotation curves as Milgrom did, we require
\begin{eqnarray}
\label{FentropicFmilgrom}
&& F_{entropic} \approx m \frac{a^2}{2\,a_0} = F_{Milgrom} \approx m \sqrt{a_N a_c},
\\
\label{aMilgrom}
&& \Leftrightarrow ~ a = \left(\,4\, a_N \,a_0^2 \,a_c \, \right)^{\frac14}
=\left(\,2\, a_N \,a_0^3 / \pi \, \right)^{\frac14}.
%\,\textrm{for}\,a \ll a_0.
%&&  F_{entropic}= m \frac{a^2}{2\,a_0} = F_{Milgrom}= m \sqrt{a_N a_0} \\
%&& \Rightarrow a = \left(\,4\, a_N \,a_0^3 \, \right)^{\frac14},
%~~ \textrm{if}~~ a \ll a_0 .
\end{eqnarray}
Numerically, it turns out that $2 \pi a_c \approx a_0$, and so we set $a_c = a_0/(2\pi )$ for simplicity.
To reproduce the flat rotation curves, we first need to realize that
$
F_{centripetal} = m \frac{a^2}{2\,a_0}
$\,
for $a \ll a_0$. Thus, the terminal velocity $v$ should be determined from \,
$
m \frac{a^2}{2\,a_0} = \frac{m \,v^2}{r},
$
with $a$ \,given by Eq. \eqref{aMilgrom}. Obviously, this leads to a constant $v$ (independent of $r$)
and hence the flat rotation curves.
%a
%= \left(\,4\, a_N \,a_0^2 \,a_c \, \right)^{\frac14}
%$=\left(\,\frac{2}{\pi}\, a_N \,a_0^3 \, \right)^{\frac14}$,
%$a = \left(\,4\, a_N \,a_0^3 \, \right)^{\frac14}$,
%which leads to flat rotation curves.

%Next, we consider the analog of the second part of Verlinde's argument.
%To get the expression for $a_N$, Verlinde starts by writing
%$a_N= 2 \pi k_B T$ with $T=2 E/ N k_B = 2 M (G_N/A)/k_B $.

On the other hand, similar to Verlinde's holographic approach which invokes the imaginary holographic
screen of radius $r$, we can write
\begin{equation}
2 \pi k_B \tilde{T} \\
= 2\pi k_B \left(\,\frac{2 \tilde{E}} {N k_B}\,\right) \\
= 4\pi \left(\,\frac{\tilde{M}} {A / G}\,\right) \\
= \frac{G\,\tilde{M}}{r^2},
\end{equation}
where $\tilde{M}$ represents the \emph{total} mass enclosed within the volume
$V = 4 \pi r^3 / 3$.\,
%Now, the question is:
What is $\tilde{M}$?
Suppose we set $\tilde{M} = M$, which means that there is only ordinary
matter enclosed by Verlinde's imaginary
holographic screen. In that case, we will have $F_{entropic}= m a_N$ even for
$a \ll a_0$. But this implies that there
is neither dark matter nor consistency with modified gravity given
by Eq. \eqref{FentropicFmilgrom} and Eq. \eqref{aMilgrom}, and as such is
obviously
incompatible with observations. The only way to be consistent with the observational data
is to have $\tilde{M} = M + M'$ where $M'$ is some unknown mass --- that is, dark
matter. Thus, we need the concept of dark matter for consistency.
%Now, to connect everything and at the same time be consistent,

In what follows, we {\it propose} that
\begin{eqnarray}
\label{Mdark}
M'
%= 2\,\left(\,\frac{a_0\, a_c}{a^2}\,\right)\, M
= \frac{1}{\pi}\,\left(\,\frac{a_0}{a}\,\right)^2\, M .
\end{eqnarray}
Note that the above formula can be generalized to AdS space, in which
case the missing matter makes a {\it negative} contribution. With Eq. \eqref{Mdark}, we can write
%$ F_{entropic}= m\,\frac{G\,M}{r^2}
%\left[\,1+\frac{1}{\pi}\,\left(\,\frac{a_0}{a}\,\right)^2\,\right].$
\begin{equation}
\label{modG}
F_{entropic} = m [\sqrt{a^2+a_0^2}-a_0] = m\,a_N
\left[\,1+\frac{1}{\pi}\,\left(\,\frac{a_0}{a}\,\right)^2\,\right]
%F_{entropic}= m\,\frac{G\,M}{r^2}
%\left[\,1+2\,\left(\,\frac{a_0}{a}\,\right)^2\,\right] ,
\end{equation}
For $a \gg a_0$, we have $ F_{entropic} \approx m a \approx m a_N$, and hence $a=a_N$.
%On the contrary,
But, for $a \ll a_0$,
we have $ F_{entropic} \approx  m \frac{a^2}{2\,a_0} \approx m a_N (1/\pi) (a_0/a)^2$.
Solving for $a$, we get $a= \left(\,2\, a_N \,a_0^3 / \pi \, \right)^{\frac14}$, which is exactly the same
expression as required
%in Eq. \eqref{aMilgrom}
for the explanation of the galactic rotation curves.
\emph{In conclusion, using the proposal as given by Eq. \eqref{Mdark}, we can actually derive MOND.}
%Notice that for $a \ll a_0$, the entropic force goes like $ F_{entropic}
%\approx  m \frac{a^2}{2\,a_0}$, which is already a signature of modified Newton's Law.
We also observe that $M'$ is greater for smaller $a$,
which is consistent with the observations that there is more dark matter in the
galactic halos than in the regions closer to the galactic centers.
%\begin{equation}
%F_{entropic} = m [\sqrt{a^2+a_0^2}-a_0] = m\,\frac{G\,M}{r^2}
%\left[\,1+\frac{1}{\pi}\,\left(\,\frac{a_0}{a}\,\right)^2\,\right]
%F_{entropic}= m\,\frac{G\,M}{r^2}
%\left[\,1+2\,\left(\,\frac{a_0}{a}\,\right)^2\,\right] ,
%\end{equation}

We can now realize the idea of CDM-MOND duality.
%as follows.
On one hand, we can interpret Eq. \eqref{modG} to mean that there is \emph{no} dark matter,
but that the law of gravity is modified.
On the other hand, we can rewrite
%$F_{entropic}$
it as
\begin{equation}
\label{DM}
F_{entropic}= m\,\frac{G\,(M+M')}{r^2} ,
\end{equation}
where $M'$ denotes the total mass of dark matter enclosed in the volume $V = 4 \pi r^3 / 3$,
which, by construction, is compatible with MOND.  We are thus led to the very intriguing
dark matter profile $M'=\frac{1}{\pi}\,\left(\,\frac{a_0}{a}\,\right)^2 \, M$.
Dark matter of this kind can behave \emph{as if} there is no dark matter but MOND.
Therefore, we call it ``MONDian dark matter". As a remark, to obtain $M'$ as a function of $r$,
one can solve the cubic equation (see Eq. \eqref{modG})
$\sqrt{a^2+a_0^2}-a_0  =  a_N \left[\,1+\frac{1}{\pi}\,\left(\,\frac{a_0}{a}\,\right)^2\,\right]$ for a
general solution of $\left(\,\frac{a_0}{a}\,\right)^2$ and substitute it into the expression for $M'$.

%It seems that we can now realize the idea of CDM-MOND duality, if we
%can claim that there is a scale at which the CDM interpretation turns into MOND
%(and vice versa).
%The price we need to pay is the relation between missing (dark) matter $M'$,
%the cosmological constant $\Lambda$ (or $a_0$) and the baryonic matter $M$,
%given by $M'=\frac{1}{\pi}\,\left(\,\frac{a_0}{a}\,\right)^2 \, M$.

\emph{Friedmann's Equations.} One important issue regarding our theory is to ensure that it
is completely compatible with cosmology. Thus, we would like to derive the corresponding
Friedmann's equations within our framework. Our derivation follows the procedure
of \cite{fe}. The FRW metric is given by
$
ds^2=-dt^2+ R(t) (dr^2+r^2\,d\Omega^2),
$
where $R(t)$ is the scale factor.
Assume that the matter sources in the universe form a perfect fluid. Then, in the
rest frame of the this fluid, the energy momentum tensor is given by
$
T_{\mu\nu}=(\rho+p)u_{\mu}u_{\nu}+p\, g_{\mu\nu},
$
where $u_{\mu}=(1,\vec{0})$ is the four velocity of the fluid.
%Notice that due to the isotropy and homogeneity of the universe, both $\rho$ and $p$
%will be dependent on time only.
Now, consider Verlinde's imaginary holographic screen of comoving radius $r$.
The physical radius would be $\tilde{r} = r R(t)$.
In de Sitter space, the net temperature observed by an accelerating observer
(with acceleration $a$) is $\tilde{T}$,\,
which leads to the entropic force discussed above.
As a result, the {\it effective} acceleration $a_{\rm eff}$ of the observer is
$
a_{\rm eff} = \sqrt{a^2+a_0^2}-a_0,
$
which is also given by
$
a_{\rm eff}= -\frac{d^2 (r R(t))}{dt^2} = - \ddot{R}\, r.
$
Using
%the relation between $\tilde{T}$ and $\tilde{M}$ we have
$2 \pi k_B \tilde{T}= \frac{G\,\tilde{M}}{r^2 R^2} $,
we get $\ddot{R} = - \frac{G\,\tilde{M}}{r^3 R^2}.$
Following \cite{fe}, in a fully relativistic situation, we replace $\tilde{M}$ by the
active gravitational (Tolman-Komar) mass
$
\mathcal{M} = \frac{1}{4\pi G} \int\, dV\, R_{\mu\nu} u^{\mu} u^{\nu}.
$
By Einstein's field equation, we obtain
$
\mathcal{M} = 2 \int\, dV\, \left(\,T_{\mu\nu} - \frac{1}{2} T g_{\mu\nu}+\frac{\Lambda}{8\pi G}g_{\mu\nu}
\,\right) u^{\mu} u^{\nu} =
%$
%Given the fact that
%$
%T_{\mu\nu} u^{\mu} u^{\nu} = \rho$, and
%$T = -\rho + 3 p$, with
%$g_{\mu\nu} u^{\mu} u^{\nu} = -1
%$,
%we deduce that
%$
%\mathcal{M} =
\left(\,\frac{4}{3}\pi r^3 R^3 \,\right) \left[\,(\rho+3 p)-\frac{\Lambda}{4\pi G}\right].
$
Finally, it follows that
\begin{equation}
\frac{\ddot{R}}{R} = - \frac{4\pi G}{3} (\rho+3 p)+\frac{\Lambda}{3},
\end{equation}
%which is exactly one of Friedmann's equations.
%Another Friedmann's equation can be obtained by multiplying both sides of the above equation
%by $\dot{R} R$ and then integrating over time $t$. Using
%This equation together
which, with the continuity equation
$
\dot{\rho}+3 H (\rho+p)=0,
%~~ \Rightarrow ~~ p = -\rho -\frac{\dot{\rho}}{3 H},
$
%and the fact that
%$
%d(R^2 \rho) =  2 R \,\rho \,dR + R^2 d\rho,
%$
can be used to obtain the other Friedmann equation, viz
%namely
\begin{eqnarray}
H^2 =  \frac{8\pi G} {3}\rho + \frac{\Lambda}{3}.
\end{eqnarray}
%\begin{equation}
%$H^2 + k /R^2 =  8\pi G \rho /3 + \Lambda/3$,
%\end{equation}
%with integration constant $k = 1, 0$ or $-1$.

We thus conclude that the corresponding Friedmann's equations derived from our framework
are exactly the same as the usual ones.
The following remark is in order: The entropic approach has {\it not} replaced general
relativity as sometimes misconstrued.  To do cosmology, we still need Einstein's general
relativity.  Indeed we have just shown
%This implies
that our theory would be completely
consistent with the $\Lambda \textrm{CDM}$ model, if the quantity $\rho$ is interpreted as
the total energy density of ordinary matter and MONDian dark matter.  Furthermore, in
our approach, we have assumed a certain symmetry, but one that is consistent with the
FRW metric.  As usual, one can categorize metric perturbations based on a homogeneous and
isotropic background.  Of course, ultimately it is important to do cosmological
perturbation theory so as to compare with observations such as CMB.

The above computation suggests that one can in principle
have Einstein's gravity together with a MONDian dark matter source.
The departure from MOND happens when we replace $\tilde{M}$
with $\mathcal{M}$, i.e. when a non-relativistic source is
replaced by a fully relativistic source. In that case, Eq. \eqref{modG} is replaced by
%So, by cdm-mond duality, both cdm and mond should work well at the galactic scale. But
%then, why cdm works well at the cluster scale while
%mond doesn't? The reason is as follows. In a relativistic system, we need to replace
%$\tilde{M}$ by the
%active gravitational mass $\mathcal{M}$
$\sqrt{a^2+a_0^2}-a_0  = \frac{G\,\mathcal{M}}{\tilde{r}^2}$, where $\tilde{r} = rR(t)$
is the physical radius, i.e.,
\begin{equation}
\label{aCluster}
\sqrt{a^2+a_0^2}-a_0  = \frac{G\, (\,M(t)+M'(t)\,)}{\tilde{r}^2} + 4 \pi G \,p\,
\tilde{r} -\frac{\Lambda}{3}\,\tilde{r}.
\end{equation}
%where $M+M'$ is now time-dependent.
If $M'=\frac{1}{\pi}\,\left(\,\frac{a_0}{a}\,\right)^2 \, M$ really gives
the correct profile for dark matter, then Eq. \eqref{aCluster} works
well at the cluster scale without any modification of gravity.
%\sqrt{a^2+a_0^2}-a_0= a_N(t)
%\left[\,1+\frac{1}{\pi}\,\left(\,\frac{a_0}{a(t)}\,\right)^2\,\right]+ 4 \pi G \,p\,
%\tilde{r} -\frac{\Lambda}{3}\,\tilde{r}
The above expression indicates that if we naively
use MOND at the cluster scale, we would be
missing $4 \pi G \,p\, \tilde{r} -\frac{\Lambda}{3}\,\tilde{r}$\,
%. But at the level of clusters, $4 \pi G \,p\, \tilde{r}$ and/or
%$\frac{\Lambda}{3}\,\tilde{r}$
which could be significant. This may explain why MOND doesn't work well at the cluster scale,
despite the CDM-MOND duality realized at the galactic scale.

%Our line of reasoning above indeed offers a logical possibility for the
\emph{$dS/CFT$ Correspondence.} So far, our arguments have been thermodynamic
but not microscopic, and thus the precise nature of MONDian dark matter is
still obscure. Apparently, at the galactic scale, the MONDian dark matter quanta
should be massless to realize the MOND-like behavior.
However, at the cluster and cosmological scales, they
should become massive and hence CDM-like.
%This is the most difficult part of our proposal.
In what follows, we will argue that the holographic dual picture might
shed some light on this issue.

%Returning to the relation between dark matter and MOND suggested above,
%The crucial question are: what determines the scale
%for the crossover between CDM and MOND, as
%suggested by the above logic, and what is the nature of
%this MOND-ian or MONDian dark matter?
We can at least use the general concept of
holography in de Sitter space \cite{dscft} to understand a possible CDM-MOND crossover.
First of all, any holographic formulation of MONDian bulk gravity in de Sitter space
should be able to
define a dictionary between the modified bulk gravity
theory consistent with MOND, and some {\it non-gravitational}
degrees of freedom associated with
the appropriate ``boundary'' holographic screen.
Currently, a precise holographic dictionary exists only
in asymptotic AdS space \cite{adscft}. But there have been some proposals
for such a holographic
dictionary in the dS$_4$ space \cite{dscft}.
One idea is to relate the dS fields $\phi$
to their AdS counterparts $\psi$ through a non-local
transformation \cite{dscft}:
$
\psi (Y) = \int dX K(X,Y) \phi(X),
$
where $dX$ is the invariant measure on $dS_4$ and
$K(X,Y)$ is a non-local kernel that commutes with the
isometries of dS$_4$.
As in the AdS/CFT correspondence, one computes the on-shell
bulk action $S_{bulk}$ and relates it to the appropriate boundary
correlators.
%(Essentially, one reinterprets the RG flow of the
%boundary non-gravitational theory in terms of
%bulk gravitational equations of motion, and then
%rewrites the generating functional of vacuum correlators
%of the boundary theory in terms of a semi-classical
%wave function of the bulk universe with specific boundary
%conditions.)
Therefore, given a modified bulk (dS$_4$) theory of gravity consistent with MOND,
one would expect (in the semiclassical limit) the standard holographic formula:\,
$
\langle \exp(- \int JO ) \rangle =
\exp[- S_{bulk}( \tilde{g}, \phi_{\tilde{g}},...)].
$

In fact, the holographic dual of a uniformly accelerating observer
in de Sitter space has been examined in \cite{das}. It was shown that
the Unruh formula in de Sitter space is
holographically mapped to a constant one-point function, namely
$
\langle O \rangle \sim ~\textrm{constant},
$
in a suitable coordinate system on the boundary \cite{das}.
Thus, an interpolation between
the Newtonian acceleration and MOND would
amount to modifying the one-point function in
the boundary theory. Once some non-gravitational degrees of freedom,
which are holographically dual to the bulk modified
theory of gravity consistent with MOND, are turned on,
the relevant one-point function
should get ``dressed'', and in principle, could have
different values along the renormalization group flow.
The question is: Where is the crossover between
the Newtonian and MOND regimes in the bulk?
It is reasonable to conjecture
that one can separate the
global de Sitter metric from the perturbations at smaller bulk scales,
while this would not be
possible at larger bulk scales. Presumably, this crossover should
happen around the galactic scale.
Of course, only a complete microscopic theory
could answer such a detailed question.

\emph{Phenomenological Implications.}
%By construction, our MONDian dark matter is
%emergent, in the sense that it is introduced along with the holographic dual
%description of MOND.
The usual way of determining the effective mass
of CDM particles is to first assume some
couplings to the Standard Model particles and impose
some ``parity-like'' quantum numbers that insure stable dark matter.
Then, one computes the relevant cross-sections and plugs them
into Boltzmann's kinetic equation. Finally, one compares the relic abundance
with cosmological constraints, such as the WMAP data \cite{dark}.
%We expect that
This logic should be repeatable in our case.

The MONDian dark matter
could lead to some \emph{distinctive phenomenological
implications}: \,
(1) The nature of MONDian dark matter quanta is constrained
by the holographic non-gravitational degrees of freedom dual to the bulk modified theory of
gravity. Thus, not any dark matter quantum numbers
would be allowed. For instance, vector and scalar degrees of freedom
may be preferred, as suggested by \cite{teves}.
\,(2) The couplings of MONDian dark matter to
the Standard Model particles could be
nonstandard, and perhaps even of a spin-orbit type,
as implied by remarkable particle physics realizations
of the Unruh effect \cite{higuchi}.
%This would be a distinctive feature
%of our dark matter model.
\,(3) In our proposal, the total mass of MONDian dark matter is related to the
cosmological constant as well as the total mass of ordinary matter.
This seems to suggest that the microscopic MONDian dark matter degrees of freedom
would know about the cosmological constant. Such ``non-locality" is obviously a unique
feature not shared by any other dark matter candidates.
\,(4) Our scheme (see Eq. (6)) may also hint at a fixed energy density ratio between the different
cosmological components of the Universe, thus helping to alleviate the coincidence problem.

\emph{Conclusions.} In this note, we have provided a
holographic dual description of Milgrom's scaling
associated with galactic rotation curves. We have proposed
a duality between cold dark matter and
modified Newtonian dynamics (MOND), encapsulated in the
new concept of MONDian dark matter.
Work on a phenomenological model of MONDian dark matter is in progress. \\

%We conclude our Letter with the following speculation.
%It is amazing that the spin-polarization of relativistic elementary particles in
%a magnetic field is related to the Unruh effect \cite{higuchi}.
%We think that this is not a coincidence and that a viable phenomenological
%model of MONDian dark matter might be constructed by using such a deep
%connection.

{\bf Acknowledgments:}
We wish to thank Nahum Arav, Patrick Huber, Mike Kavic, Raju Raghavan, John Simonetti,
Tatsu Takeuchi, Chia Tze, Sourish Dutta, Robert Scherrer and Leo Piilonen for useful
conversations.
%and Leo Piilonen for asking us about Verlinde's paper.
CMH, DM, and YJN are supported in part by DOE
under contract
DE-FG05-85ER40226,
DE-FG05-92ER40677 and DE-FG02-06ER41418 respectively.

\end{document}